\documentclass[twocolumn,showpacs,preprintnumbers,amsmath,amssymb,prl]{revtex4}
\usepackage{color}

\usepackage{graphicx}
\usepackage{dcolumn}
\usepackage{bm}
\usepackage{subfigure}
\usepackage{amssymb}

\begin{document}

\title{Is diffusion anomalous in two-dimensional Yukawa liquids?}

\author{T. Ott}
\affiliation{%
    Christian-Albrechts-Universit\"at zu Kiel, Institut f\"ur Theoretische Physik und Astrophysik, Leibnizstra\ss{}e 15, 24098 Kiel, Germany
}%

\author{M. Bonitz}%
\affiliation{%
    Christian-Albrechts-Universit\"at zu Kiel, Institut f\"ur Theoretische Physik und Astrophysik, Leibnizstra\ss{}e 15, 24098 Kiel, Germany
}%

\date{\today}

\begin{abstract}
There have recently been many predictions of ``superdiffusion'' in two-dimensional strongly coupled Yukawa systems, both by computer simulations and in dusty plasma experiments, with substantially varying diffusion exponents. Here we show that the results crucially depend on the strength of dissipation and the time instant of the measurement. For sufficiently large friction even subdiffusion is possible. 
However, there are strong indications that, in the long-time limit, anomalous diffusion vanishes for dissipative as well as for frictionless systems. 

\end{abstract}

\maketitle


Two-dimensional (2D) systems are important models for a large variety of processes on liquid or solid surfaces as well as for low-dimensional structures in condensed matter. It is long known that reduction of the dimensionality to two is accompanied by a number of anomalies in the collective properties of many-body systems including Bose condensation and superfluidity, the quantum Hall effect or phase transitions (Kosterlitz-Thouless scenario). Even purely classical systems have been predicted to exhibit anomalies in 2D which are seen e.g. in fundamental transport properties such as diffusion \cite{alder70}. 

Among classical 2D systems, monolayers of particles interacting via a Coulomb or Yukawa potential 
which can be produced in ion traps, e.g. \cite{ion-traps} and dusty plasmas \cite{review} constitute a particularly interesting generic example which allows to study 2D anomalies with unprecedented accuracy. It is, therefore, not surprising that diffusion processes in 2D Yukawa liquids (2DYL) have, over the last decade, attracted considerable interest, both in theoretical works~\cite{Vaulina2006,Liu2007,Liu2008,Ott2008,Hou2009,Ott2009a,Donko2009} and experiments in dusty plasmas~\cite{Nunomura2006,Juan1998,Juan2001,Lai2002,Quinn2002,Ratynskaia2005,Ratynskaia2006,Liu2008a,Liu2008,Chan2009}. 
While some authors found no peculiarities \cite{Vaulina2006,Nunomura2006}, the majority of these works has reported significant deviations from ``normal'' diffusion, i.e. from Fick's law and Einstein's formula for the mean-squared displacement (MSD), $u_r \sim t^{\alpha}$ with $\alpha=1$. There were many observations of {\em superdiffusion} ($\alpha>1$), i.e. enhanced diffusion \cite{Liu2007, Liu2008,Ott2008, Hou2009, Ott2009a, Donko2009,Juan1998,Juan2001,Lai2002,Quinn2002,Ratynskaia2005,Ratynskaia2006,Liu2008a}, but in some cases also of reduced diffusion, i.e. {\em subdiffusion} ($\alpha<1$)~\cite{Liu2007, Hou2009}. The strength of the diffusion can be conveniently quantified by the value of the exponent $\alpha$. However, there is a substantial scatter of the reported experimental and theoretical data, ranging from below $1.0$ to values as high as $1.3$. The origin of these differences is unknown, and even the existence of anomalous diffusion is being debated. 

The theoretical investigations were based on equilibrium computer simulations and have concentrated on the idealized 2D as well as on the quasi-2D case \cite{Ott2008}, both with and without dissipation included. Possible sources of deviations lie in differences in the complex plasma conditions such as screening strength $\kappa$, coupling parameter $\Gamma$ and neutral gas friction. In dusty plasma experiments \cite{review}, moreover, various forms of friction are always present as are additional energy sources arising from the plasma discharge giving rise to a nonequilibrium driven-dissipative system.


The aim of this Letter is to resolve the puzzle about the character of diffusion in 2DYL. To this end, we systematically analyze the effects of the interaction range (screening), coupling (temperature) and dissipation. Furthermore, it is well known that the diffusion exponent $\alpha$ varies in time: initially it equals $2$ (free ballistic motion) and lateron it is expected to approach some fixed value which determines whether the system exhibits normal, sub- or superdiffusion. It has been noted ~\cite{Liu2008a,Donko2009,Ott2009a,Hou2009a,Chan2009} that this value depends on the choice of the time window during which it is recorded. This may be another explanation for the large scatter of $\alpha$ values reported previously but raises the question about the correct procedure. We therefore, explore the whole time dependence of $\alpha$ in detail extending the simulations to very long times. The results can be summarized as follows: the character of diffusion depends on the magnitude of dissipation. While for weak friction, superdiffusion is observed, an increase of friction eventually gives rise to normal diffusion and subdiffusion. However, this turns out to be only a transient phenomenon. For 
sufficiently large observation times, diffusion transforms to normal diffusion.

{\bf Model and simulation idea.}
Our 2DYL consists of $N$ particles contained in a quadratic monolayer with periodic boundary conditions. The particle motion is modelled by coupled Langevin equations, 
\begin{equation}
    m_i \ddot{{\vec r}}_i =\vec F_i - m_i \bar \nu \vec v_i + \vec y_i \, \qquad i=1\dots N\, \label{eq:langevin}
\end{equation}
where $\bar \nu$ is the friction coefficient and $\vec F_i$ the Yukawa force,
\[
   \vec F_i = -\frac{Q^2}{4\pi\varepsilon_0} \sum_{j\neq i}\left ( \nabla \frac{e^{-r/\lambda_D}}{r} \right )\Bigg \vert_{r=\vert \vec r_i - \vec r_j \vert} 
\]
with the Debye screening length $\lambda_D$ and charge $Q$. 
$\vec y_i(t)$ is a Gaussian white noise with zero mean and the standard deviation
$
   \langle y_{\alpha,i}(t_0)y_{\beta,j}(t_0+t)\rangle=2k_BT\bar \nu\delta_{ij}\delta_{\alpha\beta}\,\delta(t)
$,
where $\alpha,\beta\in \{x,y\}$ and $T$ is the temperature.
We choose as the unit of length the Wigner-Seitz radius $a_{ws}=(n\pi)^{-1/2}$, where $n$ is the areal density, and as the unit of time the inverse of the plasma frequency $\omega_p = (Q^2/2\pi\varepsilon_0ma_{ws}^3)^{1/2}$. In thermodynamic equilibrium, the system is thus fully described by three parameters -- the dimensionless inverse screening length $\kappa = a_{ws}/\lambda_D$, the friction coefficient $\nu = \bar\nu/\omega_p$ and the Coulomb coupling parameter $\Gamma=Q^2/(4\pi\varepsilon_0 ak_BT)$.
Eq. \eqref{eq:langevin} is solved by standard Langevin dynamics~\cite{Allen1987} up to a  maximum observation time $t_{obs}$ which is limited by the condition that no collective oscillations, e.g., sound waves, should be able to traverse the entire simulation box of length $L$ during the measurement \cite{Donko2009}. Thus, $t_{obs} < L/v_s$ should be satisfied where $v_s$ is the (sound) velocity of the fastest mode which for 2DYL is well known~\cite{Peeters1987,Kalman2004,Donko2008}. By simulating large systems with $100,000$ particles, we achieve $t_{obs}=2,500$ plasma cycles, 
extending previous results~\cite{Liu2007, Ott2008,Hou2009,Ott2009a}
 by one order of magnitude. 

{\bf Mean squared displacement.} In examining the character of diffusion, the most important quantity is the MSD, $u_r(t)=\langle \vert \vec r(t) - \vec r(t_0) \vert^2 \rangle_N$, where the averaging is over all particles. A general parametrization is
 \begin{equation}
  u_r(t)=D_0t^{\alpha(t)}\, , \label{eq:msd-t}
 \end{equation}
where in case of normal diffusion, at long times $\alpha$ approaches unity, giving rise to the diffusion coefficient $D=u_r(t)/4t$. In contrast, in the case of anomalous diffusion, $\alpha$ differs from unity:  $\alpha>1$ ($\alpha<1$) is associated with superdiffusion (subdiffusion).


\begin{figure}
\includegraphics[scale=0.7]{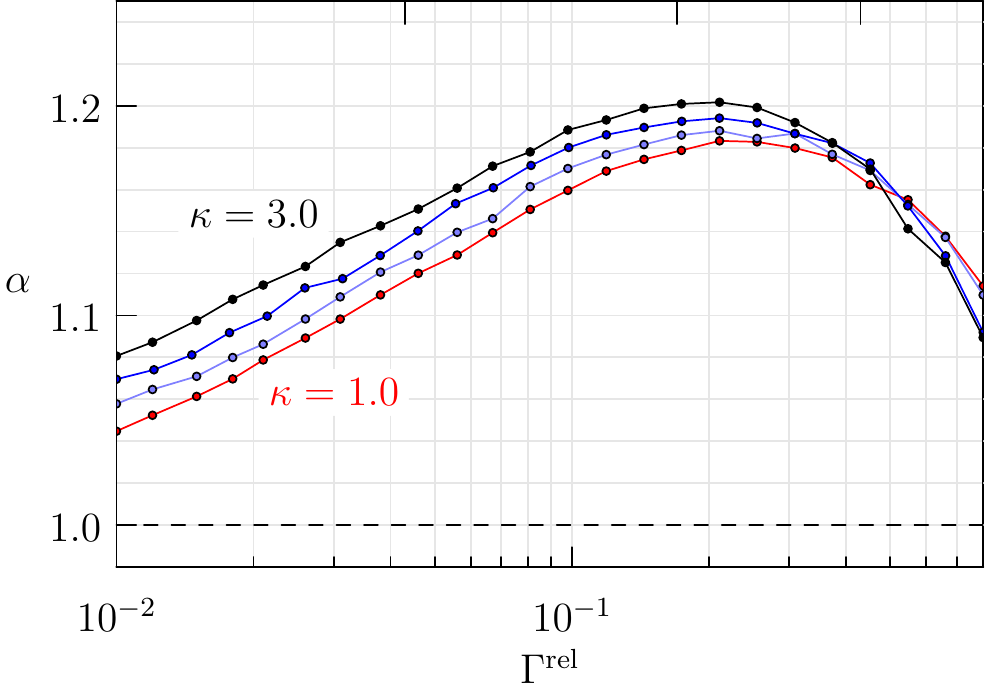}
\caption{\label{fig:kxgamma}(Color online) Diffusion exponent $\alpha$ as a function of $\Gamma^\textrm{rel}$ for different 
values of $\kappa$ ($1.0, 2.0, 2.5, 3$) obtained from a best fit of Eq.~\eqref{eq:msd-t} in the time window $\omega_pt\in [100,320]$
 The three ticks on the upper x-axis indicate the three values of $\Gamma^\textrm{rel}$ 
used in Figs.~\ref{fig:t-alpha}--\ref{fig:gamma_alpha}. 
}
\end{figure}

In a first series of simulations with fewer particles and zero friction, we establish the overall diffusion trends in 2DYL, i.e., the dependence of $\alpha$ on both the coupling strength and the range $1/\kappa$ of the pair interaction. To be able to compare the data for different $\kappa$, we introduce the relative coupling $\Gamma^\textrm{rel}$ by normalizing the Coulomb coupling parameter to the crystallization point $\Gamma_c(\kappa)$, i.e., $\Gamma^\textrm{rel}=\Gamma/\Gamma_c$~\cite{Ott2009a}. This is only one possible definition of the physical coupling -- see e.g. Refs.~\cite{Hartmann2005,Kalman2004,Vaulina2002a}  for other definitions -- but a particularly physically intuitive one. 
The results of these simulations are shown in Fig.~\ref{fig:kxgamma}. A first observation is that $\alpha$ is larger than unity for a large range of $\Gamma^\textrm{rel}$, clearly indicating superdiffusion confirming earlier simulations~\cite{Liu2007,Hou2009, Ott2009a}. Our data are significantly more comprehensive covering the full range of coupling strengths from the gas phase to the strongly coupled liquid. The dependence is non-monotonic with a maximum around $\Gamma^\textrm{rel}=0.2$. 
Our data show that superdiffusion is strongly dependent on the degree of correlations in the system. For weak coupling, the system is dominated by binary interactions, and $\alpha$ is only slightly larger than unity. When the coupling is increased, superdiffusion becomes more pronounced because collective effects grow. After a maximum, the value of $\alpha$ is again decreasing since particle movement is increasingly hindered by entrapment in local potential minima (``cages'') and the onset of crystallization.
A second conclusion from Fig.~\ref{fig:kxgamma} is the weak but systematic dependence of $\alpha$ on $\kappa$ in the range $\kappa=1\dots 3$. This trend has been first observed in \cite{Ott2009a} for three values of $\Gamma^\textrm{rel}$ and is here confirmed for the whole range of coupling strengths.

{\bf Effect of dissipation.} We now turn to the influence of friction and to the long-time behavior of diffusion. To achieve long simulation times, $t_{obs}$, it is advantageous to consider large $\kappa$ values since the sound speed $v_s$ diminishes rapidly with $\kappa$. Below we will, therefore, concentrate 
on the case $\kappa=3$ which -- due to the weak $\kappa$ dependence observed in Fig.~\ref{fig:kxgamma} -- is representative for a 2DYL. This allows us to perform extensive large simulations with $N=100,000$ up to $t\omega_p=2,500$.

\begin{figure}
\includegraphics[scale=0.55]{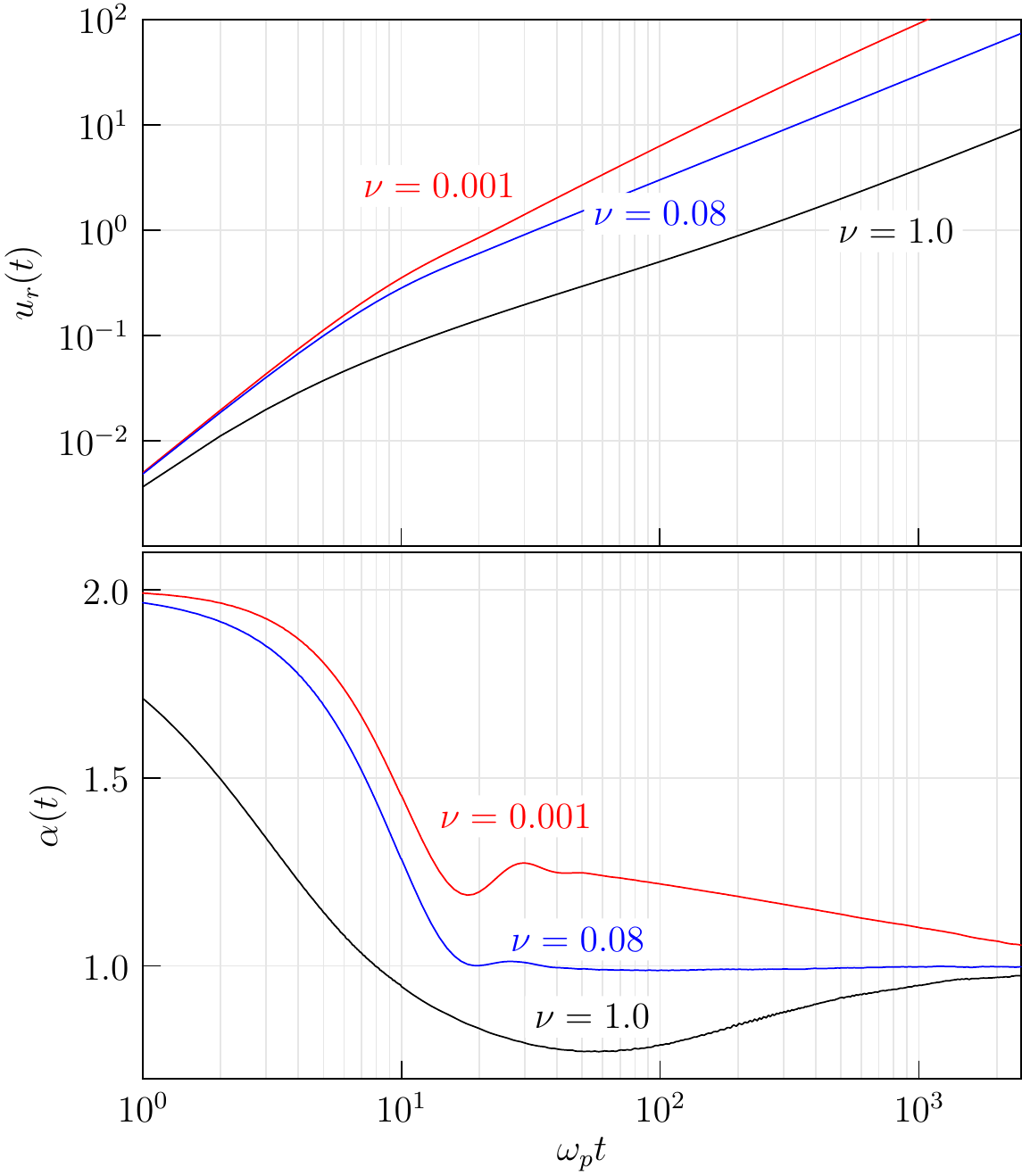}
\caption{\label{fig:t-alpha}(Color online) (a)~$u_r(t)$ for $\Gamma^\textrm{rel}=0.17$ and 
$\nu=0.001, 0.08, 1.0$. 
(b) Corresponding $\alpha(t)$
}
\end{figure}

In Fig.~\ref{fig:t-alpha}(a), we show $u_r(t)$ for three values of the friction coefficient $\nu$ at a fixed coupling $\Gamma^\textrm{rel}=0.17$ close to the maximum of the curves in Fig.~\ref{fig:kxgamma}, corresponding to a moderately coupled liquid state. The ballistic regime, $\omega_pt\lesssim 5$, is followed by a transition period after which the MSD eventually seems to approach an asymptotic behavior characterized by different slopes for different $\nu$. 
%
From these curves it is clear that the common method to extract $\alpha$ from the average slope of the MSD in a given time window, e.g., \cite{Liu2008a, Ott2008, Hou2009}, can be rather ambiguous, depending on the width and position of the latter. Therefore, here we avoid any averaging and, instead, study an \emph{instantaneous} diffusion exponent $\alpha^\textrm{inst}$ as the derivative $\alpha^\textrm{inst}(t)=\partial_{\ln t} \ln[u_r(\ln t)]$.

Fig.~\ref{fig:t-alpha}(b) shows $\alpha^\textrm{inst}(t)$ for the same curves (we drop the superscript in the following). 
Evidently, the instantaneous exponent $\alpha(t)$ is much more sensitive to the time-dependence of the diffusive motion which turns out to be non-trivial. Consider first the curve for the friction value $\nu=0.08$. Here, $\alpha(t)$, 
after a transient period of about $20$ plasma periods, approaches its asymptotic value $\alpha=1$, i.e. the system exhibits normal diffusion. In contrast, for the low-friction case, $\nu = 0.001$, after a transient of similar length $\alpha(t)$ remains significantly larger than unity corresponding to superdiffusive behavior. However, no constant value $\alpha(t)$ is reached within our simulation time, rather $\alpha(t)$ continues to decrease monotonically towards unity.
Even more interesting is the behavior at large friction, $\nu=1$: after the ballistic regime, a broad intermediate phase is observed where the motion changes 
from superdiffusion to subdiffusion. The latter persist for $10\lesssim \omega_pt \lesssim 2000$ after which the diffusion is normal (Here, and in the following, we use the threshold $\alpha=1.03$ [$\alpha=0.97$] to locate the boundary between normal diffusion and superdiffusion [subdiffusion]). 

\begin{figure}
\includegraphics[scale=0.63]{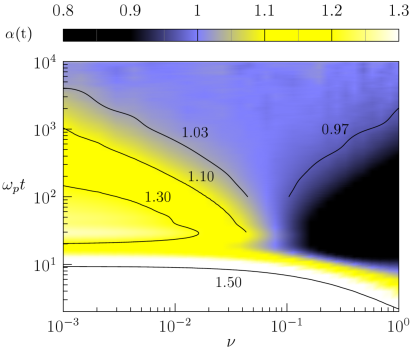}
\includegraphics[scale=0.63]{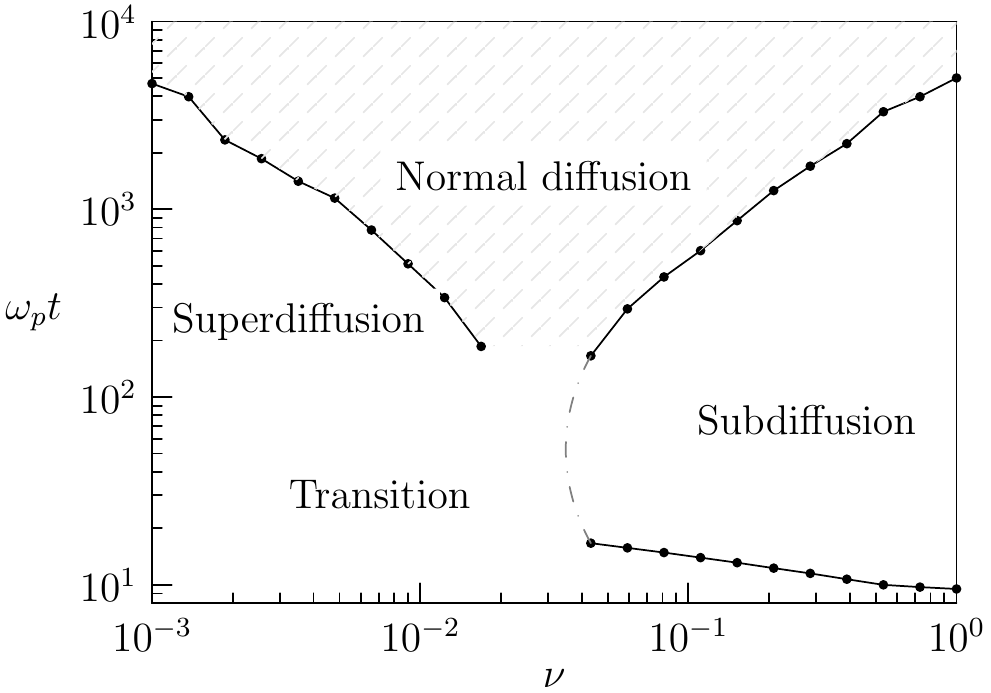}
\caption{\label{fig:nu-t-alpha1}(Color online) Instantaneous diffusion exponent $\alpha(t)$ as a function of friction coefficient $\nu$ and time $\omega_pt$ for $\Gamma^\textrm{rel}=0.17$. 
(a) Density representation with contour lines for $\alpha=1.50, 1.03, 0.97$
(b) Schematic representation with labelled regions of super-, sub- and normal diffusion. }
\end{figure}

\begin{figure}
\includegraphics[scale=0.63]{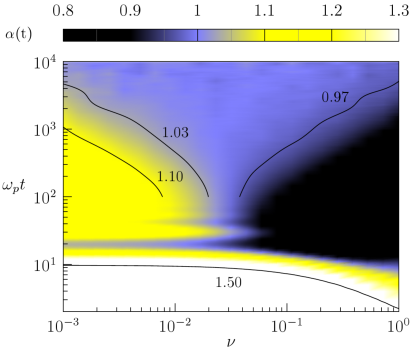}
\includegraphics[scale=0.63]{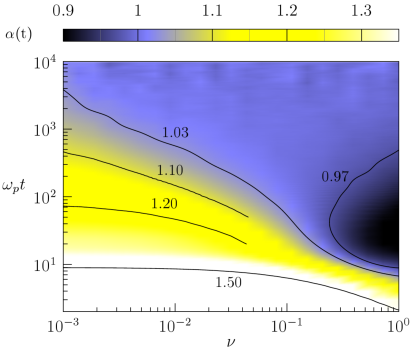}
\caption{\label{fig:nu-t-alpha2}(Color online) Instantaneous diffusion exponent $\alpha(t)$ as a function of friction coefficient $\nu$ and time $\omega_pt$ for $\Gamma^\textrm{rel}=0.43$ (top) and $\Gamma^\textrm{rel}=0.043$ (bottom). 
}
\end{figure}

{\bf Transient and long-time behavior for different couplings.}
To systematically examine the influence of dissipation on $\alpha(t)$, we repeat these simulations for a larger number of friction coefficients $\nu$ and different couplings. The entire time dependence of $\alpha(t)$ for arbitrary $\nu$ at $\Gamma^\textrm{rel}=0.17$ is comprised in Fig.~\ref{fig:nu-t-alpha1}. A first observation is that, in all cases, dissipation ultimately induces a transition to normal diffusion. Superdiffusion may exist, but it is a transient effect which is possible only at sufficiently low friction. The time window of superdiffusion rapidly decreases as dissipation grows. At high friction, instead, transient subdiffusion is observed within a time window which is growing with increasing friction. Only within a relatively small friction interval a direct crossover from ballistic motion to normal diffusion is observed. 
This is an interesting special case where the combined effect of friction and thermal fluctuations prevents the build-up of collective motions in the liquid.
%
%
%
The corresponding graphs for the limits of very strong and very weak coupling, $\Gamma^\textrm{rel}=0.43$ and $0.043$, respectively, are shown in Figs. \ref{fig:nu-t-alpha2}. Here, the overall picture is the same as for $\Gamma^\textrm{rel}=0.17$ with the main modification that, for lower [stronger] coupling, the region of transient superdiffusivity is extended towards higher [lower] friction, while the subdiffusive region is diminished [increased]. Thus we conclude that we have established the general trends of time-dependent diffusion processes in dissipative 2DYL.

\begin{figure}[h]
\includegraphics[scale=0.65]{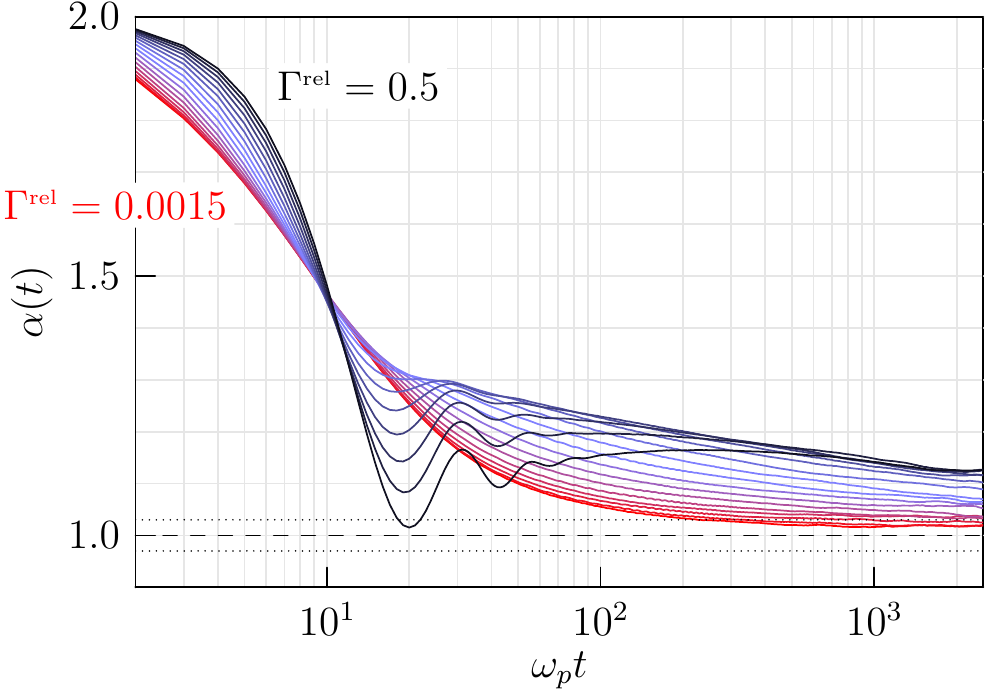}
\caption{\label{fig:gamma_alpha}(Color online) Instantaneous diffusion exponent $\alpha(t)$ 
for $\nu=0$. 
Subsequent curves differ in $\Gamma^\textrm{rel}$ by a factor of $1.41$.
}
\end{figure}

Now the interesting question arises whether the emergence of normal diffusion at very long time scales is the sole consequence of dissipation or if this is an intrinsic property of 2DYL which is observed also in the friction-less limit. To this end, we have performed 
an additional series of simulations for $\nu=0$ and a broad range of couplings the results of which are displayed in Fig.~\ref{fig:gamma_alpha}. For high couplings, $\alpha(t)$ shows oscillations during the transient regime with a frequency close to the Einstein frequency~\cite{Hartmann2005} which are due to caged motion of the particles. This transient regime is followed by a long period of superdiffusive motion which seems to tend 
towards normal diffusion for longer times. 
On the other hand, at weak coupling, $\alpha(t)$ does not exhibit oscillations but rather decays monotonically from $2$ to $1$. Whether in the long time limit normal diffusion will emerge, cannot be answered definitely from our data due to the limited observation time. However, an extrapolation of the data seems to indicate that $\alpha(t)$ converges towards unity within two more time decades, i.e. for $t\gtrsim 10^5 \omega_{pl}^{-1}$, for all coupling strengths. 


In summary, in this Letter we have presented a comprehensive numerical analysis of diffusion in 2DYL, both 
with and without friction. We have clarified the effects of coupling, interaction range and dissipation and, most importantly, the time dependence of the diffusion exponent over long time scales. Our results explain the partly contradicting results of earlier studies and confirm the observation of superdiffusive behavior. 
However, this turns out to be a transient effect existing only in a finite time window and below a critical  dissipation which depends on the coupling strength. For larger dissipation, instead, transient subdiffusion is observed. For the limit of non-dissipative 2DYL our simulation results provide strong indications that here  anomalous diffusion vanishes as well on sufficiently long time scales $\omega_pt\gtrsim 10^5$. 
We underline that our results do not rule out superdiffusion in dusty plasma experiments because they are performed under stationary nonequilibrium conditions. To verify the relevance of nonequilibrium states, we suggest to perform measurements of the time dependence of $\alpha$ and compare it to our data. 

The observed nontrivial anomalous diffusion of classical particles arises from their confinement to two dimensions. Similar transient anomalies should show up in other transport properties. Moreover, it may be expected that 2D anomalies of correlated quantum systems will exhibit similar time-dependent modifications as well which 
should influence their excitation spectra.

This work is supported by the Deutsche Forschungsgemeinschaft via SFB-TR 24.


\begin{thebibliography}{23}

\bibitem{alder70} B.~Alder, and T.E.~Wainwright, Phys. Rev. A {\bf 1}, 18 (1970)

\bibitem{ion-traps} D.J.~Wineland \textit{et al.}, Phys. Rev. Lett. \textbf{59}, 2935 (1987)

\bibitem{review} for a recent overview see  P.K.~Shukla, and A.A. Mamun, {\em Introduction to Dusty Plasma Physics} (Institute of Physics, Bristol, 2002).

\expandafter\ifx\csname natexlab\endcsname\relax\def\natexlab#1{#1}\fi
\expandafter\ifx\csname bibnamefont\endcsname\relax
  \def\bibnamefont#1{#1}\fi
\expandafter\ifx\csname bibfnamefont\endcsname\relax
  \def\bibfnamefont#1{#1}\fi
\expandafter\ifx\csname citenamefont\endcsname\relax
  \def\citenamefont#1{#1}\fi
\expandafter\ifx\csname url\endcsname\relax
  \def\url#1{\texttt{#1}}\fi
\expandafter\ifx\csname urlprefix\endcsname\relax\def\urlprefix{URL }\fi
\providecommand{\bibinfo}[2]{#2}
\providecommand{\eprint}[2][]{\url{#2}}

\bibitem[{\citenamefont{Nunomura et~al.}(2006)\citenamefont{Nunomura, Samsonov,
  Zhdanov, and Morfill}}]{Nunomura2006}
\bibinfo{author}{\bibfnamefont{S.}~\bibnamefont{Nunomura}},
  \bibinfo{author}{\bibfnamefont{D.}~\bibnamefont{Samsonov}},
  \bibinfo{author}{\bibfnamefont{S.}~\bibnamefont{Zhdanov}}, \bibnamefont{and}
  \bibinfo{author}{\bibfnamefont{G.}~\bibnamefont{Morfill}},
  \bibinfo{journal}{Phys. Rev. Lett.} \textbf{\bibinfo{volume}{96}},
  \bibinfo{eid}{015003} (\bibinfo{year}{2006}).


\bibitem[{\citenamefont{Vaulina and Dranzhevski}(2006)}]{Vaulina2006}
\bibinfo{author}{\bibfnamefont{O.~S.} \bibnamefont{Vaulina}} \bibnamefont{and}
  \bibinfo{author}{\bibfnamefont{I.~E.} \bibnamefont{Dranzhevski}},
  \bibinfo{journal}{Phys. Scr.} \textbf{\bibinfo{volume}{73}},
  \bibinfo{pages}{577} (\bibinfo{year}{2006}).

\bibitem[{\citenamefont{Liu and Goree}(2007)}]{Liu2007}
\bibinfo{author}{\bibfnamefont{B.}~\bibnamefont{Liu}} \bibnamefont{and}
  \bibinfo{author}{\bibfnamefont{J.}~\bibnamefont{Goree}},
  \bibinfo{journal}{Phys. Rev. E} \textbf{\bibinfo{volume}{75}},
  \bibinfo{pages}{16405} (\bibinfo{year}{2007}).

\bibitem[{\citenamefont{Liu et~al.}(2008)\citenamefont{Liu, Goree, and
  Feng}}]{Liu2008}
\bibinfo{author}{\bibfnamefont{B.}~\bibnamefont{Liu}},
  \bibinfo{author}{\bibfnamefont{J.}~\bibnamefont{Goree}}, \bibnamefont{and}
  \bibinfo{author}{\bibfnamefont{Y.}~\bibnamefont{Feng}},
  \bibinfo{journal}{Phys. Rev. E} \textbf{\bibinfo{volume}{78}},
  \bibinfo{eid}{046403} (\bibinfo{year}{2008}).

\bibitem[{\citenamefont{Ott et~al.}(2008)\citenamefont{Ott, Bonitz, Donk{\'o},
  and Hartmann}}]{Ott2008}
\bibinfo{author}{\bibfnamefont{T.}~\bibnamefont{Ott et al.}},
  \bibinfo{journal}{Phys. Rev. E} \textbf{\bibinfo{volume}{78}},
  \bibinfo{eid}{026409} (\bibinfo{year}{2008}).

\bibitem[{\citenamefont{Hou et~al.}(2009)\citenamefont{Hou, Piel, and
  Shukla}}]{Hou2009}
\bibinfo{author}{\bibfnamefont{L.-J.} \bibnamefont{Hou et al.}},
  \bibinfo{journal}{Phys. Rev. Lett.} \textbf{\bibinfo{volume}{102}},
  \bibinfo{eid}{085002} (\bibinfo{year}{2009}).

\bibitem[{\citenamefont{Ott et~al.}(2009{\natexlab{a}})\citenamefont{Ott,
  Bonitz, and Hartmann}}]{Ott2009a}
\bibinfo{author}{\bibfnamefont{T.}~\bibnamefont{Ott}},
  \bibinfo{author}{\bibfnamefont{M.}~\bibnamefont{Bonitz}}, \bibnamefont{and}
  \bibinfo{author}{\bibfnamefont{P.}~\bibnamefont{Hartmann}},
  \bibinfo{journal}{Phys. Rev. Lett.}  (\bibinfo{year}{2009}{\natexlab{a}}),
  \bibinfo{note}{accepted for publication}.

\bibitem[{\citenamefont{Donk\'{o} et~al.}(2009)\citenamefont{Donk\'{o}, Goree,
  Hartmann, and Liu}}]{Donko2009}
\bibinfo{author}{\bibfnamefont{Z.}~\bibnamefont{Donk\'{o} et al.}},
  \bibinfo{journal}{Phys. Rev. E} \textbf{\bibinfo{volume}{79}},
  \bibinfo{eid}{026401} (\bibinfo{year}{2009}).

\bibitem[{\citenamefont{Chan et~al.}(2009)\citenamefont{Chan, Io, and
  Lin}}]{Chan2009}
\bibinfo{author}{\bibfnamefont{C.}~\bibnamefont{Chan}},
  \bibinfo{author}{\bibfnamefont{C.}~\bibnamefont{Io}}, \bibnamefont{and}
  \bibinfo{author}{\bibfnamefont{I.}~\bibnamefont{Lin}},
  \bibinfo{journal}{Contrib. Plasma Phys.} \textbf{\bibinfo{volume}{49}},
  \bibinfo{pages}{215} (\bibinfo{year}{2009}).

\bibitem[{\citenamefont{Juan and I}(1998)}]{Juan1998}
\bibinfo{author}{\bibfnamefont{W.-T.} \bibnamefont{Juan}} \bibnamefont{and}
  \bibinfo{author}{\bibfnamefont{L.}~\bibnamefont{I}}, \bibinfo{journal}{Phys.
  Rev. Lett.} \textbf{\bibinfo{volume}{80}}, \bibinfo{pages}{3073}
  (\bibinfo{year}{1998}).

\bibitem[{\citenamefont{Juan et~al.}(2001)\citenamefont{Juan, Chen, and
  I}}]{Juan2001}
\bibinfo{author}{\bibfnamefont{W.-T.} \bibnamefont{Juan}},
  \bibinfo{author}{\bibfnamefont{M.-H.} \bibnamefont{Chen}}, \bibnamefont{and}
  \bibinfo{author}{\bibfnamefont{L.}~\bibnamefont{I}}, \bibinfo{journal}{Phys.
  Rev. E} \textbf{\bibinfo{volume}{64}}, \bibinfo{pages}{016402}
  (\bibinfo{year}{2001}).

\bibitem[{\citenamefont{Lai and I}(2002)}]{Lai2002}
\bibinfo{author}{\bibfnamefont{Y.-J.} \bibnamefont{Lai}} \bibnamefont{and}
  \bibinfo{author}{\bibfnamefont{L.}~\bibnamefont{I}}, \bibinfo{journal}{Phys.
  Rev. Lett.} \textbf{\bibinfo{volume}{89}}, \bibinfo{pages}{155002}
  (\bibinfo{year}{2002}).

\bibitem[{\citenamefont{Quinn and Goree}(2002)}]{Quinn2002}
\bibinfo{author}{\bibfnamefont{R.}~\bibnamefont{Quinn}} \bibnamefont{and}
  \bibinfo{author}{\bibfnamefont{J.}~\bibnamefont{Goree}},
  \bibinfo{journal}{Phys. Rev. Lett.} \textbf{\bibinfo{volume}{88}},
  \bibinfo{pages}{195001} (\bibinfo{year}{2002}).

\bibitem[{\citenamefont{Ratynskaia et~al.}(2005)\citenamefont{Ratynskaia,
  Knapek, Rypdal, Khrapak, and Morfill}}]{Ratynskaia2005}
\bibinfo{author}{\bibfnamefont{S.}~\bibnamefont{Ratynskaia}},
  \bibinfo{author}{\bibfnamefont{C.}~\bibnamefont{Knapek}},
  \bibinfo{author}{\bibfnamefont{K.}~\bibnamefont{Rypdal}},
  \bibinfo{author}{\bibfnamefont{S.}~\bibnamefont{Khrapak}}, \bibnamefont{and}
  \bibinfo{author}{\bibfnamefont{G.}~\bibnamefont{Morfill}},
  \bibinfo{journal}{Phys. Plasmas} \textbf{\bibinfo{volume}{12}},
  \bibinfo{eid}{022302} (\bibinfo{year}{2005}).

\bibitem[{\citenamefont{Ratynskaia et~al.}(2006)\citenamefont{Ratynskaia,
  Rypdal, Knapek, Khrapak, Milovanov, Ivlev, Rasmussen, and
  Morfill}}]{Ratynskaia2006}
\bibinfo{author}{\bibfnamefont{S.}~\bibnamefont{Ratynskaia et al.}},
   \bibinfo{journal}{Phys. Rev. Lett.}
  \textbf{\bibinfo{volume}{96}}, \bibinfo{eid}{105010} (\bibinfo{year}{2006}).

\bibitem[{\citenamefont{Liu and Goree}(2008)}]{Liu2008a}
\bibinfo{author}{\bibfnamefont{B.}~\bibnamefont{Liu}} \bibnamefont{and}
  \bibinfo{author}{\bibfnamefont{J.}~\bibnamefont{Goree}},
  \bibinfo{journal}{Phys. Rev. Lett.} \textbf{\bibinfo{volume}{100}},
  \bibinfo{eid}{055003} (\bibinfo{year}{2008}).

\bibitem[{\citenamefont{Allen and Tildesley}(1987)}]{Allen1987}
\bibinfo{author}{\bibfnamefont{M.}~\bibnamefont{Allen}} \bibnamefont{and}
  \bibinfo{author}{\bibfnamefont{D.}~\bibnamefont{Tildesley}},
  \emph{\bibinfo{title}{{Computer Simulation of Liquids}}}
  (\bibinfo{publisher}{Clarendon Press}, \bibinfo{year}{1987}).


\bibitem[{\citenamefont{Peeters and Wu}(1987)}]{Peeters1987}
\bibinfo{author}{\bibfnamefont{F.~M.} \bibnamefont{Peeters}} \bibnamefont{and}
  \bibinfo{author}{\bibfnamefont{X.}~\bibnamefont{Wu}}, \bibinfo{journal}{Phys.
  Rev. A} \textbf{\bibinfo{volume}{35}}, \bibinfo{pages}{3109}
  (\bibinfo{year}{1987}).

\bibitem[{\citenamefont{Kalman et~al.}(2004)
\citenamefont{Kalman, Hartmann, Donk{\'o}, and Rosenberg}}]{Kalman2004}
\bibinfo{author}{\bibfnamefont{G.}~\bibnamefont{Kalman et al.}},
  \bibinfo{journal}{Phys. Rev. Lett.} \textbf{\bibinfo{volume}{92}},
  \bibinfo{pages}{65001} (\bibinfo{year}{2004}).

\bibitem[{\citenamefont{Donko et~al.}(2008)\citenamefont{Donko, Kalman, and
  Hartmann}}]{Donko2008}
\bibinfo{author}{\bibfnamefont{Z.}~\bibnamefont{Donko}},
  \bibinfo{author}{\bibfnamefont{G.~J.} \bibnamefont{Kalman}},
  \bibnamefont{and} \bibinfo{author}{\bibfnamefont{P.}~\bibnamefont{Hartmann}},
  \bibinfo{journal}{J. Phys. Condens. Matter} \textbf{\bibinfo{volume}{20}},
  \bibinfo{pages}{413101} (\bibinfo{year}{2008}).

\bibitem[{\citenamefont{Hartmann et~al.}(2005)\citenamefont{Hartmann, Kalman,
  Donk{\'o}, and Kutasi}}]{Hartmann2005}
  \bibinfo{author}{\bibfnamefont{P.}~\bibnamefont{Hartmann et al.}},
  \bibinfo{journal}{Phys. Rev. E} \textbf{\bibinfo{volume}{72}},
  \bibinfo{eid}{026409} (\bibinfo{year}{2005}).

\bibitem[{\citenamefont{Vaulina et~al.}(2002)\citenamefont{Vaulina, Khrapak,
  and Morfill}}]{Vaulina2002a}
\bibinfo{author}{\bibfnamefont{O.}~\bibnamefont{Vaulina et al.}},
  \bibinfo{journal}{Phys. Rev. E} \textbf{\bibinfo{volume}{66}},
  \bibinfo{pages}{016404} (\bibinfo{year}{2002}).

\bibitem[{\citenamefont{{Hou} et~al.}(2009)\citenamefont{{Hou}, {Piel}, and
  {Shukla}}}]{Hou2009a}
\bibinfo{author}{\bibfnamefont{L.~J.} \bibnamefont{{Hou}}},
  \bibinfo{author}{\bibfnamefont{A.}~\bibnamefont{{Piel}}}, \bibnamefont{and}
  \bibinfo{author}{\bibfnamefont{P.~K.} \bibnamefont{{Shukla}}},
  \bibinfo{journal}{Phys. Rev. Lett.}  (\bibinfo{year}{2009}),
  \bibinfo{note}{accepted for publication}.


\end{thebibliography}

\end{document}